\documentstyle[preprint,prb,aps,psfig]{revtex}

\begin{document}
\draft

\title{Multifractal analysis of the metal-insulator transition in
  anisotropic systems}

\author{Frank Milde, Rudolf A. R\"{o}mer, and Michael Schreiber}
\address{Institut f\"{u}r Physik, Technische Universit\"{a}t, D-09107
  Chemnitz, Federal Republic of Germany}

\date{Version: September 26, 1996; printed \today} \maketitle

\begin{abstract}
  We study the Anderson model of localization with anisotropic hopping
  in three dimensions for weakly coupled chains and weakly coupled
  planes. The eigenstates of the Hamiltonian, as computed by Lanczos
  diagonalization for systems of sizes up to $48^3$, show multifractal
  behavior at the metal-insulator transition even for strong
  anisotropy.  The critical disorder strength $W_c$ determined from
  the system size dependence of the singularity spectra is in a
  reasonable agreement with a recent study using transfer matrix
  methods. But the respective spectrum at $W_c$ deviates from the
  ``characteristic spectrum'' determined for the isotropic system.
  This indicates a quantitative difference of the multifractal
  properties of states of the anisotropic as compared to the isotropic
  system.  Further, we calculate the Kubo conductivity for given
  anisotropies by exact diagonalization.  Already for small system
  sizes of only $12^3$ sites we observe a rapidly decreasing
  conductivity in the directions with reduced hopping if the coupling
  becomes weaker.
\end{abstract}


\narrowtext

%
%

\section{Introduction}

It is well known that the three dimensional (3D) isotropic Anderson
model exhibits a metal-insulator transition (MIT): Increasing the
disorder of the random potential site energies causes the wave
functions to localize.  \cite{And} There exists a mobility edge in the
energy-disorder diagram which separates extended from localized
eigenstates.  In order to determine these critical disorders $W_c(E)$
accurately, the transfer-matrix method (TMM) together with the
one-parameter finite-size scaling hypothesis applied to quasi-1D bars
has been used with much success in the past.
\cite{Kramer_Kinnon83,Kramer_Kinnon90,Kramer_Kinnon93} Recently, the
{\em anisotropic} Anderson model has received much attention in
connection with the anisotropic transport properties of the high $T_c$
cuprates and a possible contradiction to the scaling theory of
localization was mentioned\cite{Rojo} supported by a diagrammatic
analysis.  \cite{Abrikosov} However, recent TMM
studies\cite{AEbenen1,AEbenen2,AKetten} show that the one-parameter
scaling theory is still valid and further that an MIT exists even for
strong hopping anisotropy $\gamma$.  The values of the critical
disorder in the band center were found to follow a power law
$W_c\propto(1-\gamma)^\beta$ independent of the orientation of the
quasi-1D bar. $\beta$ was argued to be independent of the strength of
the anisotropy.

Here, we shall study the problem of Anderson localization by a
different method: we focus our attention directly on the
eigenfunctions of the Hamiltonian.  In an infinite system the wave
functions are expected to be localized on the insulating side and
extended on the metallic side even arbitrarily close to the
transition.  As first suggested by Aoki\cite{Aoki83} the fractal
nature of the critical eigenstates can connect these discrepant
characteristics.  Indeed, large fluctuations of the wave functions
have been observed numerically which dominate --- at least at small
length scales --- the character of the states and invalidate the
simple notions of exponentially localized or homogeneously extended
states.  Approaching the transition these fluctuations increase and at
the critical disorder they are expected to occur on all length scales.
It has been shown\cite{GS91} that such wave functions are multifractal
entities. To characterize the eigenstates of the isotropic Anderson
model the singularity spectrum $f(\alpha)$ has been used
successfully.\cite{GS91} A {\em characteristic spectrum} $f_c(\alpha)$
was shown to determine the mobility edge independent of the
microscopic details of the sample.\cite{GS92} Further, around its
maximum, $f_c(\alpha)$ agrees well with an analytical result of
Wegner\cite{Wegner} based on a nonlinear $\sigma$ model calculation.
Near the critical disorder $W_c$, characteristic changes of
$f(\alpha)$ were observed when the system size was
increased.\cite{GS93} These distinguish the localized and the extended
character of the states and therefore allow us to determine the
transition directly from multifractal properties of eigenstates.

It is our aim in the present work to use and extend these concepts for
the case of anisotropic hopping. In Sec.~\ref{sec-modell} we introduce
our notation and define the anisotropies of weakly coupled planes and
weakly coupled chains. We next recall the concepts and methods of the
multifractal analysis employed in the sequel.  Using the hypothesis of
a characteristic singularity spectrum, we estimate the critical
disorders in Sec.~\ref{sec-res-us}.  To check the validity of the
hypothesis we analyze the system size dependence of the multifractal
properties and compare our results with TMM
data.\cite{AEbenen2,AKetten} For completeness, we also study the
conductivity of small samples of anisotropic systems in
Sec.~\ref{sec-cond}.  We discuss our results in Sec.~\ref{sec-concl}.

%
%

\section{The Anderson model with anisotropic hopping}
\label{sec-modell}

The Anderson Hamiltonian is given as\cite{And}
\begin{equation}
  \label{Hand}
  H = \sum_{i} \epsilon_{i} | i \rangle\langle i | + \sum_{i \ne j}
  t_{ij} | i \rangle\langle j | \quad .
\end{equation}
Here, the sites $i=(x,y,z)$ form a regular cubic lattice of size $N^3$
and the potential site energies $\epsilon_{i}$ are as usual taken to
be randomly distributed in the interval $[-W/2,+W/2]$.  The transfer
integrals $t_{ij}$ are restricted to nearest neighbors and depend only
on the spatial direction, so $t_{ij}$ can either be $t_x$, $t_y$ or
$t_z$.  We set the energy scale by normalizing the largest $t_{ij}$ to
$1$.

Following Ref.~\onlinecite{AEbenen2}, we study two possibilities of
anisotropic transport: (i) {\em weakly coupled planes} with
$t_x=t_y=1$, $t_z=1-\gamma$ and (ii) {\em weakly coupled chains} with
$t_x=t_y=1-\gamma$, $t_z=1$. Here the parameter $\gamma\in [0,1]$
describes the strength of the anisotropy. Hence, for $\gamma=0$ we
recover the isotropic 3D case and $\gamma=1$ corresponds to $N$
independent planes or $N^2$ independent chains. The direction with
normal (reduced) transfer integral is called the parallel
(perpendicular) direction.

The Lanczos algorithm,\cite{Cullum} which is well suited for the
diagonalization of sparse matrices, allows us to solve the eigenvalue
equation $H \Psi(E)=E \Psi(E)$ for system sizes up to $N=72$, yielding
eigenvalue/eigenvector pairs in a requested energy range.  We use
state-of-the-art workstations and a parallel computer with 128 PowerPC
processors.  It takes about 11 hours to diagonalize the Anderson
Hamiltonian with $N=48$ on the parallel machine using 8 processors.
The workstations need about 35 hours for the same calculation.  Since
we also have to perform a statistical averaging over different
disorder configurations, we have restricted the systematic
investigations to sizes up to $N=48$. In order to allow a direct
comparison with the results of
Refs.~\onlinecite{AEbenen1,AEbenen2,AKetten}, we restrict our study to
the states in the center of the band such that $E=0$. Numerically this
is the worst case because of the high density of states there which
requires a very large tridiagonal matrix in the Lanczos algorithm to
determine the eigenvalues.

%
%

\section{Multifractal analysis}
\label{sec-multi}

Fractal measures are widely used in physics to characterize objects
such as percolating clusters, random walks, and random surfaces.
\cite{mandel,Bunde,feder} The common geometric feature of such point
sets is the self-similarity: Parts of the set are similar to the
whole, at least in a statistical sense.  However, for fluctuating
physical quantities such as the probability amplitude of an eigenstate
$\Psi(E)$ of the Anderson model, the appropriate concept is given by
the multifractal measures: If the mentioned fluctuations are
statistically the same on every length scale, i.e., if all the
moments of the investigated quantity are self similar, the object is
(statistically) self-affine and is called a multifractal.

A characteristic property of multifractals is their singularity
spectrum $f(\alpha)$.\cite{feder} Let us briefly describe an algorithm
to determine this quantity, based on the standard box counting
procedure: We consider a volume $L^D$ in our $D$ dimensional space
which contains the support of the physical variable, i.e., all points
where the variable is defined. We cover it with a number of ''boxes''
of linear size $r=\delta L$. The actual shape of the boxes is not
important, they may be spheres as well.  Next, we determine the
contents $\mu_i(\delta)$ of each box $i$ by summing or integrating the
investigated quantity over the part of the support inside the box.
For a self-affine object one finds a power-law dependence
$\mu_i(\delta)\propto\delta^{\alpha_i}$ in the limit $\delta\to 0$.
The so-defined singularity strength $\alpha_i$ is assigned to
each point of the support. The subset $S_\alpha$ which contains all
points with the same value of $\alpha$ is a fractal with fractal
dimension $f(\alpha)$ defined by $K(\alpha,\delta) \propto
\delta^{-f(\alpha)}$.  Here, $K(\alpha,\delta)$ is the number of boxes
which cover $S_\alpha$.  A multifractal object consists of a
(infinite) number of subsets $S_\alpha$ with different fractal
dimensions.

In the present work we shall use an equivalent but numerically more
convenient algorithm to compute the singularity spectrum.  Our
physical quantity is again the probability amplitude of eigenstates.
Considering the normalized $q$th moments of the box probability
$\mu_i(q,\delta)=\mu_i^q(\delta) / \sum_k \mu_k^q(\delta)$ it is
possible to find\cite{Chhabra} a parametric expression of $f(\alpha)$
such that
\begin{equation}
  \label{f_von_alpha}
  \begin{array}{l}
    \alpha(q)=\lim_{\delta \to 0} \sum_i \mu_i(q,\delta) {\rm ln}
    \mu_i(1,\delta) / {\rm ln} \delta \quad , \\ f(q)=\lim_{\delta \to
      0} \sum_i \mu_i(q,\delta) {\rm ln} \mu_i(q,\delta) / {\rm ln
      \delta}\quad .
  \end{array}
\end{equation}
We plot the sums in (\ref{f_von_alpha}) versus ${\rm ln} \delta$ and
observe multifractal behavior if and only if the data may be
well fitted by straight lines. The slope from the linear regression
procedure used in the fit gives $f$ and $\alpha$.  Note, that a check
of the linearity is important, since the numerical procedure gives an
$f(\alpha)$ curve for nearly every distribution of the physical
variable, but without the mentioned linearity it does not indicate
multifractality.

In general, $f(\alpha)$ is a nonnegative, convex function with
$0<\alpha_{min}\leq\alpha\leq\alpha_{max}<\infty$.  The maximum of
$f(\alpha)$ at $\alpha(q=0)\equiv\alpha_0$ equals the dimension of the
support, i.e., the fractal dimension $D_f$ of the subset of points
where the investigated quantity is not zero.  For our wave functions
$D_f=D=3$ because they are nowhere exactly zero.  The whole
$f(\alpha)$ curve is below the bisector $f(\alpha)=\alpha$ except at
$\alpha(q=1)\equiv\alpha_1$ where both curves touch.  For $q=1$ the
relation $f(\alpha_1)=\alpha_1$ is fulfilled. $\alpha_1$ equals the
entropy dimension or information dimension and one can show that the
corresponding set $S_{\alpha_1}$ contains the entire
measure.\cite{feder}

\begin{figure}[p]
  \centerline{\psfig{figure=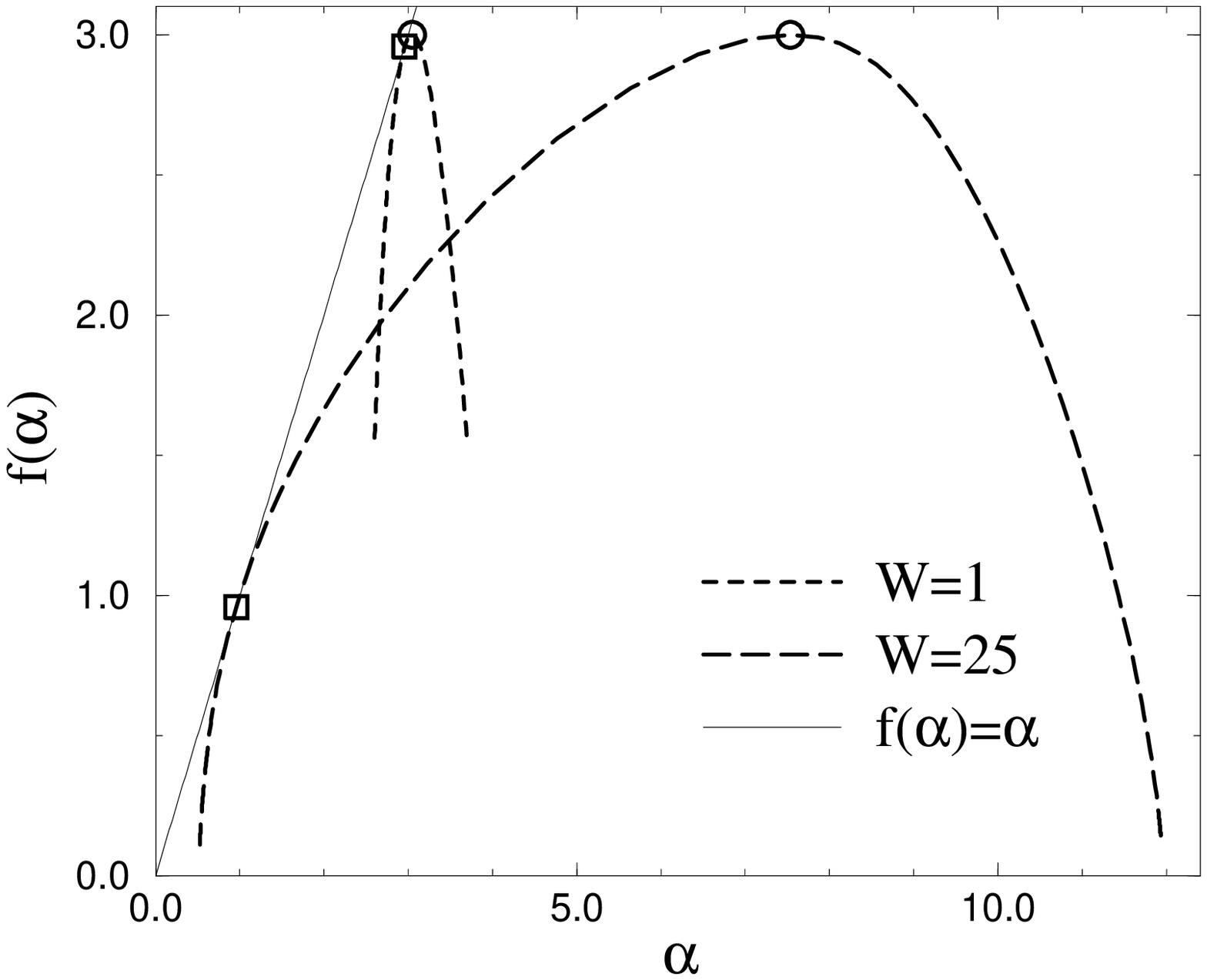}}\vspace{1cm}
  \caption{Singularity spectra of a localized ($W=1$) and an extended
    ($W=25$) state of an isotropic system with $N=48$. The circles
    ($\circ$) mark $f(\alpha_0)$ and the squares ($\Box$) mark
    $f(\alpha_1)$.}
  \label{fig:typ_spec}
\end{figure}

There are two limits which will be important for the later
interpretation of our results.  Consider a {\it D}-dimensional
support.  (i) A uniform distribution is represented by the single
point $f(\alpha=D)=D$ in the singularity spectrum, because
$\mu_i(\delta)\propto (L \delta)^D$ for every point of the support.
(ii) With increasing localization the spectrum becomes wider and an
extremely localized distribution with measure $1$ at one point and $0$
elsewhere has a spectrum which consists of two points only:
$f(\alpha=0)=0$ and $f(\alpha=\infty)=D$.  This is because the box
around the maximum has contents $1$ for each $\delta$, so $\alpha$ is
$0$ for this single point while all other points have
$\mu_i\propto\delta^{\infty}=0$.  In Fig.~\ref{fig:typ_spec} we show
two typical singularity spectra of 3D wave functions corresponding to
a localized and an extended wave function. The tendency towards the 2
limiting cases can be seen for these two examples already: The
extended state has a narrow $f(\alpha)$ curve close to $f(3)=3$ while
the localized wave function is represented by a very wide spectrum
with larger $\alpha_0$ and smaller $\alpha_1$.

%
%

\section{Calculation of critical disorders $W_c(\gamma)$}
\label{sec-res}

\subsection{Existence of multifractal eigenstates}
\label{sec-res-mf}

\begin{figure}[p]
  \centerline{\psfig{figure=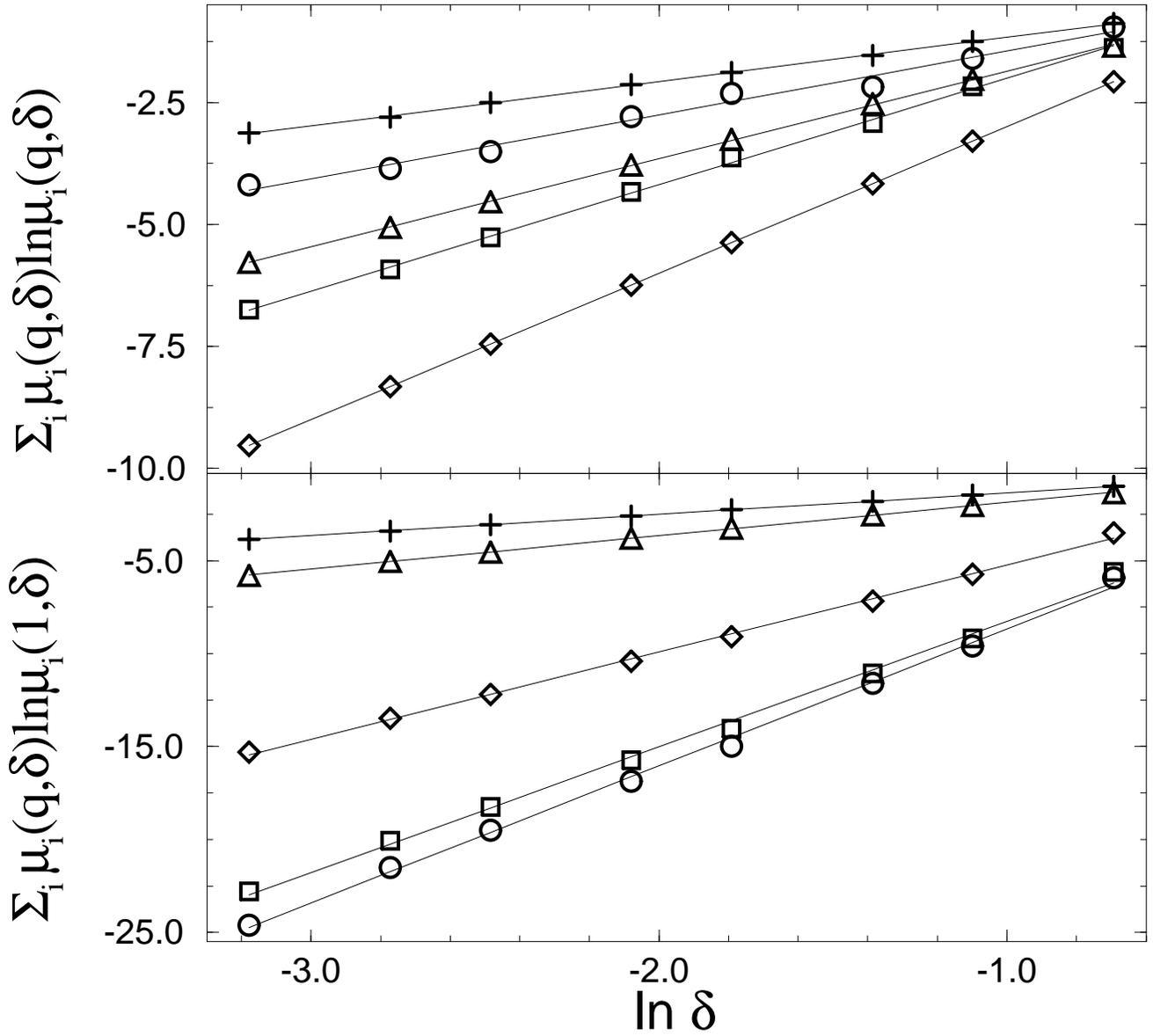}}\vspace{1cm}
  \caption{Linear regression data for the evaluation of
    Eq.~(\ref{f_von_alpha}) determining $\alpha(q)$ and $f(q)$ for
    weakly coupled planes with $\gamma=0.9$, $N=48$, $W=9$ and
    $q=-2(\circ),-1(\Box),0(\Diamond),1(\triangle),2(+)$.}
  \label{fig:lin_reg}
\end{figure}

As has been shown in Refs.~\onlinecite{AEbenen2,AKetten} by the TMM,
the anisotropic Anderson model still exhibits a MIT for all $\gamma>0$
in the band center $E=0$ and, by the general arguments given above, we
expect the wave functions at the transition point to be multifractals
just as in the isotropic case.  As a check we have computed various
eigenstates close to the proposed\cite{AEbenen2,AKetten} critical
disorders $W_c$ for system sizes up to $N=48$.  In
Fig.~\ref{fig:lin_reg}, we show the data for the linear regression of
a typical state with $W\approx W_c$. We find even for very strong
anisotropies $\gamma=0.99$ that the sums in Eq.~(\ref{f_von_alpha}),
plotted versus ${\rm ln}\delta$ are linear.  Therefore, we do find
multifractal behavior of the wave functions close to the critical
disorder for the anisotropic Anderson model.

Every singularity spectrum is characteristic only for the particular
configuration of the site energies. But for a given set of parameters
$\{W,E,\gamma\}$ the different $f(\alpha)$ curves fluctuate around one
singularity spectrum.  In order to suppress these statistical
fluctuations we average the spectra obtained from $3$ to $8$
eigenstates close to $E=0$ for 12 realizations of the random site
energies.  The averaged spectrum is thus characteristic for the set of
parameters $\{W,E,\gamma\}$ and will be used in the next sections to
compute the critical disorder $W_c$ as a function of the anisotropy
$\gamma$.

\subsection{Estimation of $W_c$ from comparison with the characteristic
  spectrum}
\label{sec-res-us}

In the isotropic case a {\em characteristic singularity spectrum}
$f_c(\alpha)$ was found previously\cite{GS95} at all points of the
mobility edge independent of the microscopic details of the system
such as the probability distribution of the site energies.  The region
close to the maximum of $f_c(\alpha)$ is described well by an
analytical result of Wegner\cite{Wegner} from the $2+\varepsilon$
expansion of the non-linear $\sigma$ model, i.e.,
\begin{equation}
  \label{Fc(a)}
  f_c(\alpha) = D - \frac{\varepsilon}{4} \left(
    \frac{D-\alpha}{\varepsilon}+1 \right)^2 + O(\varepsilon^4)
  \stackrel{\varepsilon=1}{\approx} 3 - \frac{(4-\alpha )^2}{4}.
\end{equation}
As a hypothesis we shall now assume that this characteristic
spectrum determines the transition even in the case of anisotropic
hopping.  This hypothesis is certainly valid in the limit $\gamma \to
0$ but needs further support for larger anisotropies.

We find that for each anisotropy $\gamma$ there exists a corresponding
$W_{f_c}$ such that the eigenstates are characterized by $f_c$.
Identifying $W_c=W_{f_c}$ gives us an estimate for the $\gamma$
dependence of the critical disorder.  Note that since the singularity
spectrum should be independent of the system size at the transition
point, it is sufficient to investigate small systems. We have used
systems with $N=24$ for the results presented in this section.

\subsubsection{Weakly coupled planes}
\label{sec-res-us-wcp}

Assuming the validity of $f_c$ we find a crossover between two power
laws in the $\gamma$ dependence of the critical disorder:
$W_c=55(1-\gamma)^{0.86}$ for $\gamma\ge 0.9$ and
$W_c=16.8\,(1-\gamma)^{0.35}$ for $\gamma \le 0.9$ as can be seen in
Fig.~\ref{fig:WcAEbenen}.  This does not agree with the results of
Ref.~\onlinecite{AEbenen2}, where $\beta=0.25$ has been calculated
within the self-consistent theory of localization and where the single
power law $W_c=15.4(1-\gamma)^{0.25}$ has been deduced from the TMM
data.

\subsubsection{Weakly coupled chains}
\label{sec-res-us-wcc}

\begin{figure}[p]
  \centerline{\psfig{figure=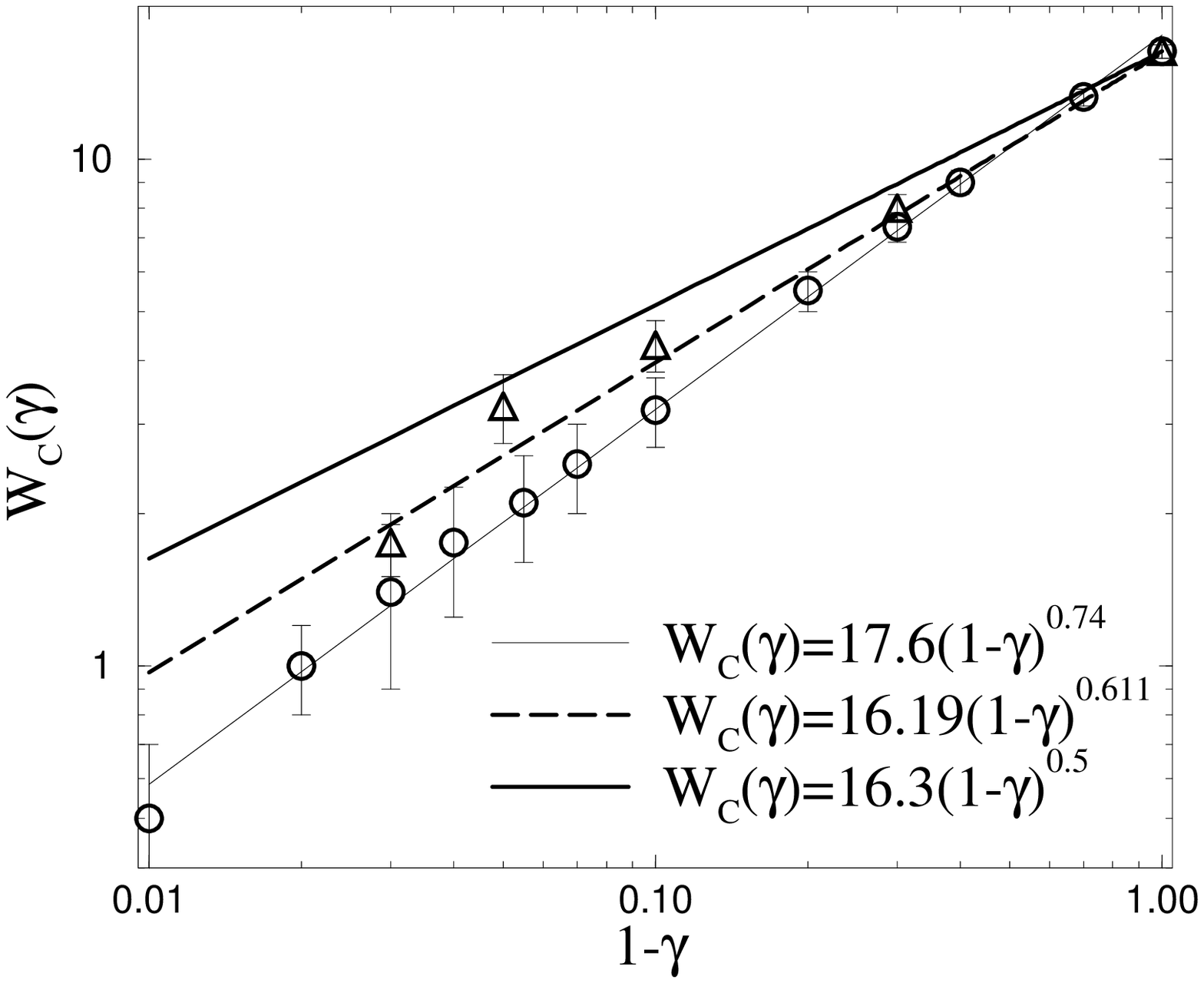}}\vspace{1cm}
  \caption{$W_c(\gamma)$ for weakly coupled chains as obtained from the
    ``characteristic spectrum'' ($\circ$) and from the system size
    dependence ($\triangle$). The thin solid line is a power law fit
    to the ($\circ$) data, the thick dashed line is the result of
    Ref.~\protect\onlinecite{AKetten}.  The thick solid line is the
    combination of the isotropic result \protect\cite{Kramer_Kinnon93}
    $W_c=16.3$ and the perturbative exponent \protect\cite{AEbenen2}
    $\beta=0.5$.}
\label{fig:WcAKetten}
\end{figure}

In Fig.~\ref{fig:WcAKetten} the results for $W_c(\gamma)$ for weakly
coupled chains are shown. Using $f_c$ we find
$W_c=17.6(1-\gamma)^{0.74}$ which is very similar to the TMM
data\cite{AKetten} $W_c=16.19 (1-\gamma)^{0.611}$.  The difference
becomes significant only for very large $\gamma\gtrsim 0.9$.  The
exponent $\beta=0.611$ was obtained from a fit of the TMM data over
the whole $\gamma$ range. For large $\gamma$ the authors of
Ref.~\onlinecite{AKetten} get $\beta=0.5$. This is consistent with the
result of Ref.~\onlinecite{AEbenen2}.

\subsection{Estimation of $W_c$ from the system-size dependence}
\label{sec-res-ssd}

We have shown in the last section, that the assumption of the
characteristic $f_c$ leads to large differences in the estimates of
$W_c$ between the TMM results \cite{AEbenen2,AKetten} and our results
based on the multifractal analysis.  Thus we will now use a more
direct method to estimate $W_c(\gamma)$ from the multifractal
properties of the eigenstates.  From the isotropic case it is
known\cite{GS92} that multifractal behavior is found not only directly
at the critical disorder $W_c$ but also close to the transition. The
reason is the finite sample size which is much smaller than the
characteristic length scales of the states close to $W_c$. In this
range the exponential decay or uniformly extended character of the
wave function is masked by large fluctuations and it is not obvious to
which side of the MIT a given state belongs.  Due to the relatively
small sample size the system is very sensitive to its boundary.
Correspondingly, a characteristic change in the singularity spectrum
is observed when the system size is increased. This change can be
evaluated to distinguish the localized or extended character of the
wave function. For an extended state the $f(\alpha)$ curve becomes
narrower and the maximum position is shifted towards smaller values of
$\alpha$, approaching the value 3. The opposite behavior is found for
a localized state.  Thus the spectra tend towards the extreme cases
discussed in Sec.~\ref{sec-multi}. Indeed we expect these limiting
cases, namely $f(3)=3$ for the metallic side and $f(0)=0$ and
$f(\infty)=3$ for the insulating side, to be reached for infinitely
large system size for any disorder except $W_c$. Only directly at the
transition the wave functions are multifractal, the fluctuations are
the same on all length scales, and $f(\alpha)$ is independent of the
system size.  This makes is feasible to determine the critical
disorder by analyzing the system-size dependence of the singularity
spectra.\cite{GS93}
\subsubsection{Weakly coupled planes}
\label{sec-res-ssd-wcp}

\begin{figure}[h]
  \centerline{\psfig{figure=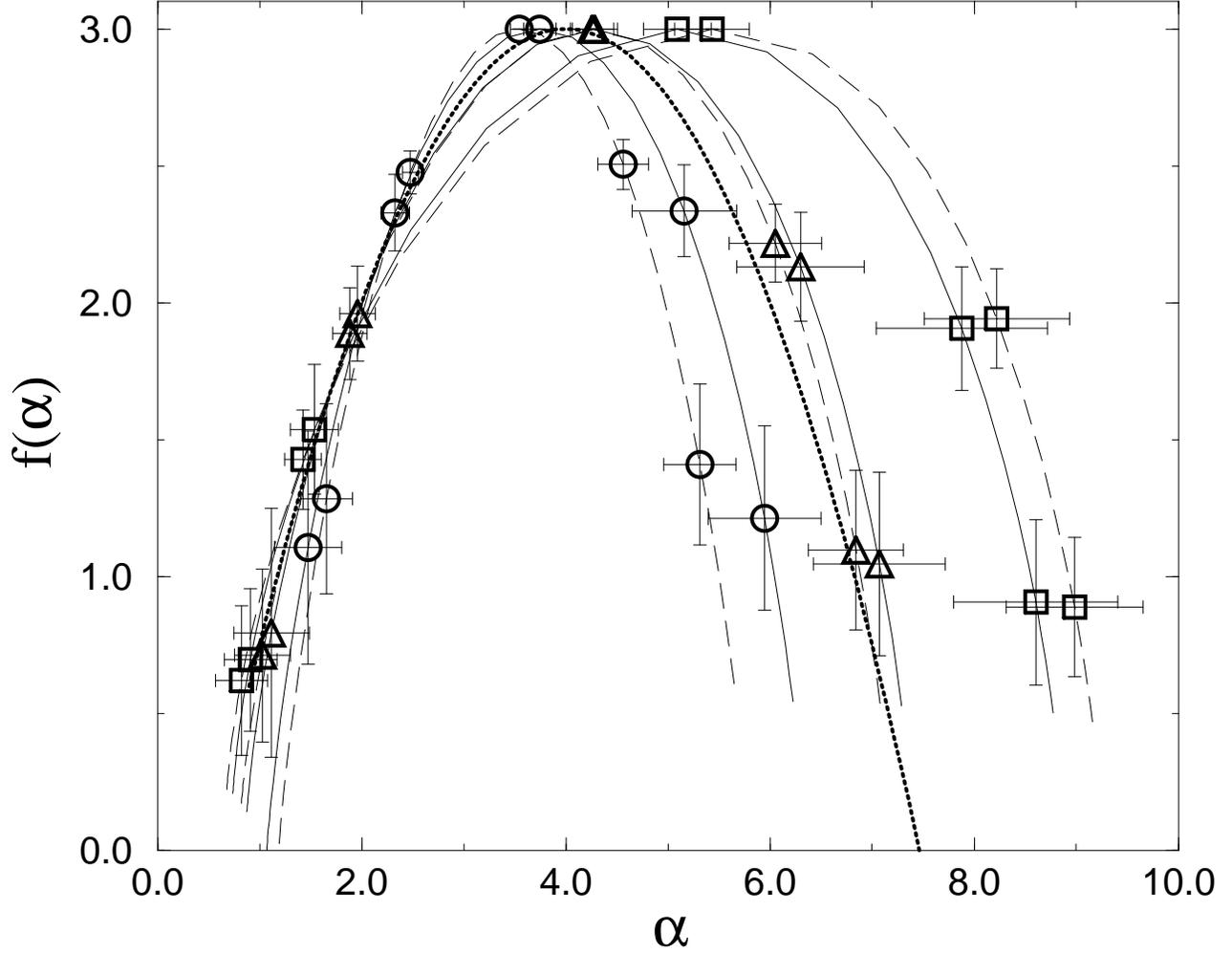}}\vspace{1cm}
  \caption{Singularity spectra for weakly coupled planes at $\gamma=0.8$
    for the two system sizes $N=18\, (-)$ and $N=42\,(-\, -)$. The
    symbols distinguish $W=8(\circ ), 10(\triangle ), 12(\Box)$ and
    indicate $\alpha(q)$ and $f(q)$ for $q=-2, -1, 0, 1, 2$ (from
    right to left). The error bars result from the linear regression
    (cp.  Fig.~\ref{fig:lin_reg}) and the average over the different
    eigenstates (cp. Sec.~\ref{sec-res-mf}).  The dotted line is the
    ''characteristic spectrum'' $f_c$.}
\label{fig:FvonW02}
\end{figure}

We show in Fig.~\ref{fig:FvonW02} an example of weakly coupled planes
with $\gamma=0.8$. The above described different behaviors of the
spectra can be seen. For $W=8$, a larger system size results in a
narrower $f(\alpha)$ curve which is characteristic for extended
states. On the other hand, the increase in the system size for $W=12$
yields a widening of the spectrum indicating localized states. The
singularity spectrum for $W=10$ is least effected by the change of the
system size and we thus conclude a critical disorder
$W_c(\gamma=0.8)=10\pm 1$. Considering the error bars, the $f(\alpha)$
curve for $W_c$ equals the characteristic spectrum $f_c$ of the
isotropic case. For moderate anisotropies $\gamma\lesssim 0.8$ this
confirms the hypothesis of a characteristic $f_c$.

\begin{figure}[h]
  \centerline{\psfig{figure=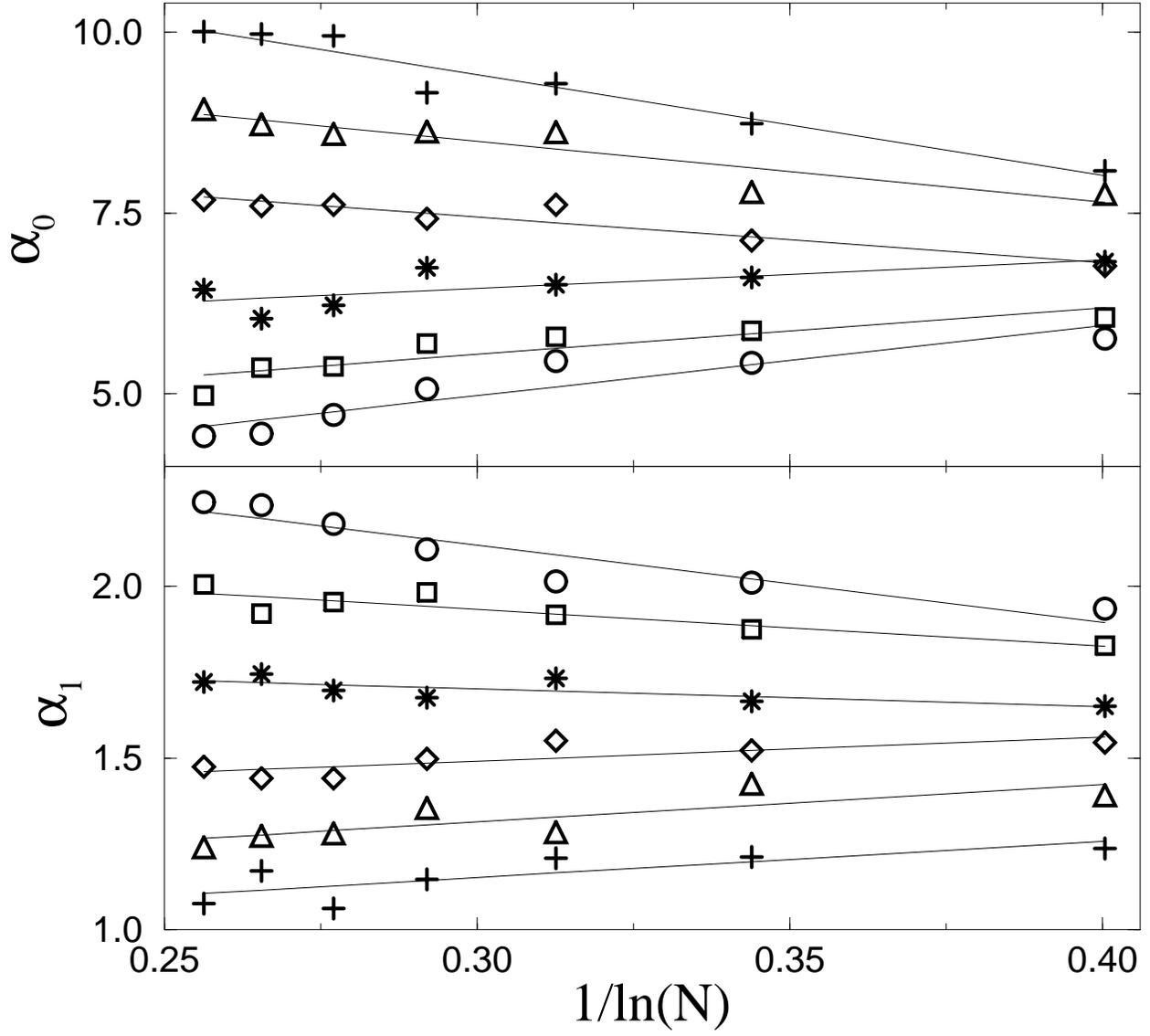}}\vspace{1cm}
  \caption{$N$ dependence of $\alpha_0$ and $\alpha_1$ for weakly coupled
    planes with $\gamma=0.96$ and $W=5.5(\circ ), 6.5(\Box),
    7.5(\star), 8.5(\diamond ), 9.5(\triangle), 10.5(+)$.}
  \label{fig:Sys04}
\end{figure}

Visual observation of the system-size dependence of the $f(\alpha)$
curves is not well suited for a systematic search for the transition.
A better method is to focus attention to special points of the spectra
such as the position $\alpha_0$ of the maximum and the information
dimension $D_1=\alpha_1$. An increase of the system size causes a
decreasing $\alpha_0$ and a increasing $\alpha_1$ for extended states
and the opposite tendency for localized states\cite{GS93,Gdiss} as
described in Sec.~\ref{sec-multi}. A constant behavior of $\alpha_0$
and $\alpha_1$ versus system size indicates $W_c$. Following
Ref.~\onlinecite{GS93} we have parameterized the system-size
dependence by $1 / {\rm ln} (N)$ which has been found to give a nearly
linear behavior of $\alpha_0$ and $\alpha_1$ thus distinguishing their
tendencies more clearly.\cite{GS93,Gdiss,Gcomm} In
Fig.~\ref{fig:Sys04} we find a constant behavior at the same value of
the disorder for both quantities and we conclude
$W_c(\gamma=0.96)=8.0\pm 0.5$.

\begin{figure}[p]
  \centerline{\psfig{figure=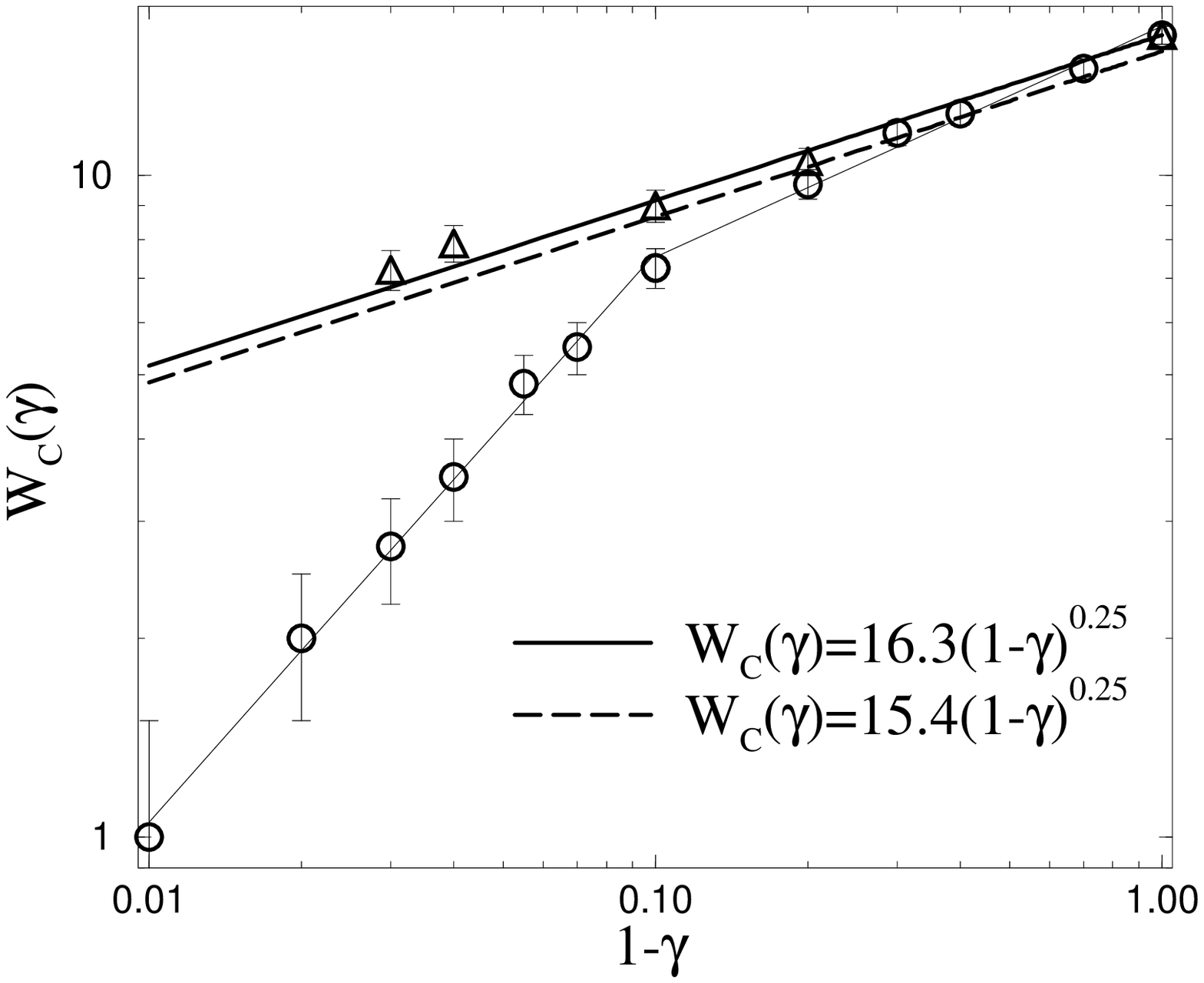}}\vspace{1cm}
  \caption{$W_c(\gamma)$ for weakly coupled planes as obtained from the
    ``characteristic spectrum'' ($\circ$) and from the system size
    dependence ($\triangle$). The thin solid lines represent the two
    power law fits to the ($\circ$) data. The thick dashed line is the
    result of Ref.~\protect\onlinecite{AEbenen2}. The thick solid line
    is a combination of the isotropic
    result\protect\cite{Kramer_Kinnon93} $W_c=16.3$ and the
    perturbative exponent\protect\cite{AEbenen2} $\beta=0.25$ which
    fits the ($\triangle$) data well.}
\label{fig:WcAEbenen}
\end{figure}

\begin{figure}[p]
  \centerline{\psfig{figure=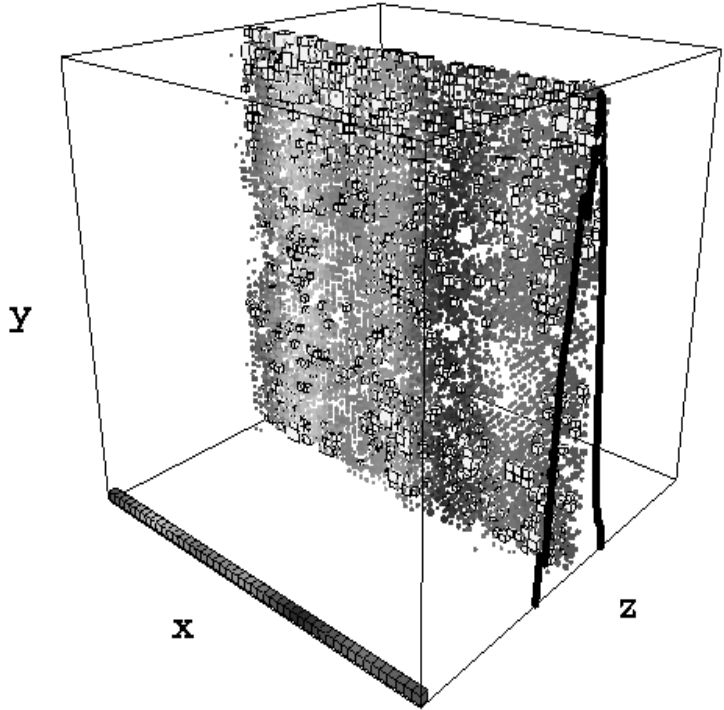,width=5in,height=7in}}
  \caption{Wave function close to the MIT for very weakly coupled planes
    with $\gamma=0.99$, $N=48$ and $W=4.5$. Every site with
    probability $|\Psi_i|^2$ larger than the average
    $N^{-3}$ is shown as a box with volume $|\Psi_i|^2 \times N$.  The
    764 cubes with $|\Psi_i|^2 \times N > \protect\sqrt{1000}$ are
    plotted in white with black edges.  The grey scale distinguishes
    between different slices of the system along the x-axis.  The
    thick solid line is the logarithm of the summed probability
    amplitude for each plane perpendicular to the z-axis. Again only
    values above $N^{-3}$ are shown.}
\label{fig:AWF01}
\end{figure}

For very weakly coupled planes we get significantly larger values of
$W_c$ than in Sec.~\ref{sec-res-us}. The new values are close to, but
slightly larger than the TMM data\cite{AEbenen2} as can be seen in
Fig.~\ref{fig:WcAEbenen}. Our data follow $W_c=16.3 (1-\gamma)^{0.25}$
which confirms the exponent $\beta=0.25$ derived
analytically.\cite{AEbenen2} We therefore conclude that $f_c$ is no
longer characteristic for the eigenstates at the MIT for weakly
coupled planes with $\gamma\gtrsim 0.8$. In our present analysis we
find wider singularity spectra which is a sign of a tendency towards
localization. An eigenstate at the transition for very strong
anisotropy $\gamma=0.99$ is shown in Fig.~\ref{fig:AWF01}. The
probability amplitude is concentrated to a few planes perpendicular to
the direction with reduced transfer. This coincides with the
observation that the localization length is smaller by a factor
$1-\gamma$ in the perpendicular direction compared with the parallel
one.\cite{AEbenen2} In the mentioned planes the wave function looks
like a fractal with holes and islands of different sizes, very similar
to the critical eigenstates of the isotropic system.  It may well be
that the cubic boxes used in the box-counting procedure for the
multifractal analysis cannot appropriately measure this fractal,
because most box sizes exceed the number of planes on which the wave
functions are concentrated. Therefore it is possible that the
deviations of $W_{f_c}$ from $W_c$ in Fig.~\ref{fig:WcAEbenen} are an
artefact of the analysis.

\subsubsection{Weakly coupled chains}
\label{sec-res-ssd-wcc}

The results for the $W_c(\gamma)$ dependence of weakly coupled chains
are shown in Fig.~\ref{fig:WcAKetten}.  They are in reasonable
agreement with the TMM data, \cite{AKetten} although we cannot
reproduce the exponent\cite{AEbenen2} $\beta=0.5$. The differences
between $W_c$ and $W_{f_c}$ are not as large as in the other case and
the multifractal properties of the critical states are therefore
similar to those of the isotropic system.

%
%

\section{Conductivity in small anisotropic systems}
\label{sec-cond}

The transport properties are determined by the localization properties
of the states. At $T=0$ localized states cannot contribute to charge
transfer and we have insulating behavior. On the other hand, extended
states yield metallic behavior. The Kubo formula following from
Fermi's golden rule provides a connection of the electrical
conductivity and the electronic states $|n\rangle$.

Let us consider an electrical AC field with frequency $\omega=\hbar E$
in $x$ direction on an sample with volume $N^3$. We assume a
half-filled band at $T=0$ such that all states with $E\leq 0$ are
occupied while all others are unoccupied. A configuration average
because of the random site energies is denoted by $\left\langle \,
\right\rangle_C$. Neglecting prefactors the real part $\sigma$ of the
conductivity is given by\cite{Greensfkt}
\begin{equation}
  \label{leit}
  \sigma(E) \sim \left\langle \sum_{nn'} |\langle n|x|n'\rangle|^2
    E_{nn'} \delta(E-E_{nn'}) \right\rangle_C \quad , E \neq 0 \quad.
\end{equation}

\begin{figure}[p]
  \centerline{\psfig{figure=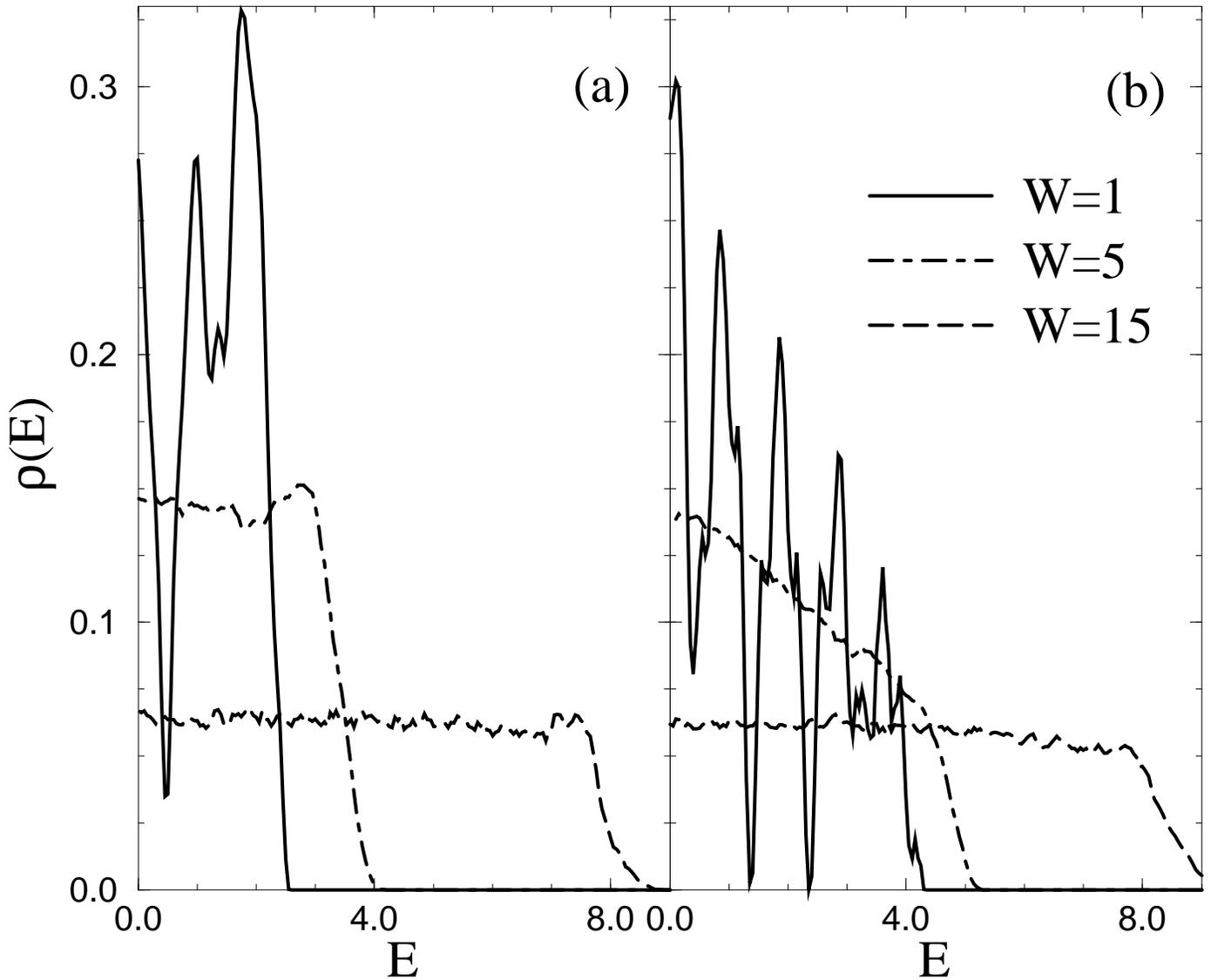}}\vspace{1cm}
  \caption{Density of states for weakly coupled chains (a) and
    planes (b) with $\gamma=0.9$ and $N=12$. The data for $W=5$ and
    $15$ agree with that of uncoupled chains or planes within the
    statistical fluctuations. The peak structure for $W=1$ is due to
    the small system size.}
  \label{fig:ne}
\end{figure}

In order to compute this quantity it is necessary to know all
eigenvalues and all eigenstates. We use the standard Householder
algorithm to diagonalize the Hamiltonian (\ref{Hand}). We average over
90 to 150 configurations to suppress the large statistical
fluctuations of $\sigma$. This limits the treatable linear system size
to $N=12$. As a consequence we encounter strong finite-size effects
for $W\leq 4$.  The small number of eigenenergies in the ordered
limit is not smeared out sufficiently by the disorder to yield a
smooth density of states $\rho(E)=\frac{1}{N^3} \left\langle \sum_n
  \delta(E-E_n) \right\rangle_C$ as shown in Fig.~\ref{fig:ne} for
$W=1$ and $\gamma=0.9$.
We also note that the density of states for larger disorder values and
$\gamma\geq 0.9$ agrees with that of the respective 1D or 2D system
within the uncertainty due to fluctuations. However, the transport
behavior of the states is completely different: In 1D and 2D there is
no MIT ($W_c=0$) while 3D systems exhibit an MIT even for very strong
anisotropy as shown in Sec.~\ref{sec-res}.

\begin{figure}[p]
  \centerline{\psfig{figure=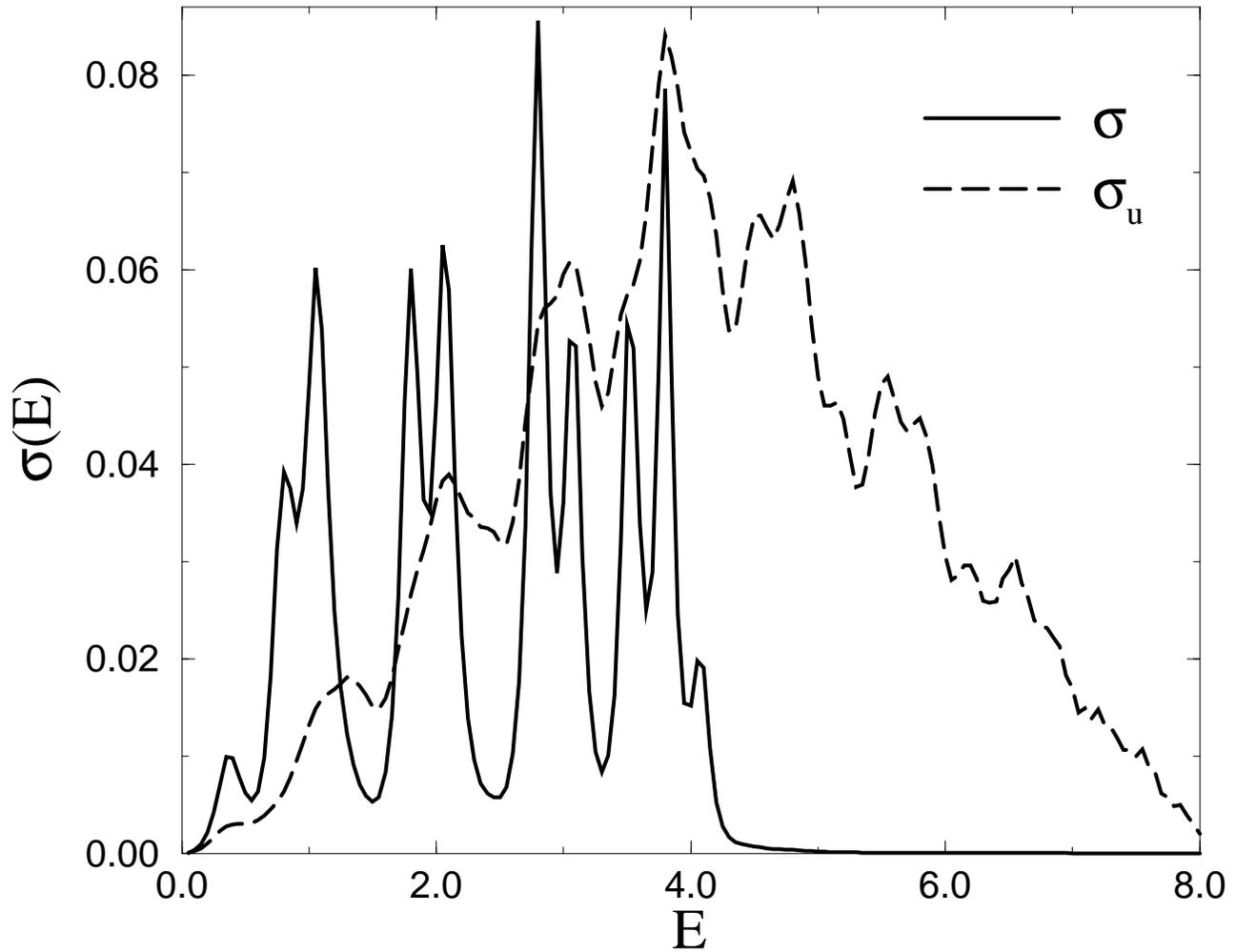}}\vspace{1cm}
  \caption{Comparison of $\sigma$ and $\sigma_u$ in weakly coupled 
    planes with $W=1$, $\gamma=0.9$, and $N=12$. $\sigma_u$ has been
    scaled to the same maximum value as $\sigma$.}
  \label{fig:CpE01+uniform}
\end{figure}

Because the characteristic length scales of the wave functions exceed
the system size it is a priori not clear whether the localization
behavior of the states has any measurable influence on the computed
conductivity.  Suppose there is no such influence, then the matrix
elements $\langle n|x|n'\rangle$ are all be equal and the conductivity
is given by the $E$-weighted joint density of states $\sigma_u(E)=
\left\langle\sum_{nn'}E_{nn'}\delta(E-E_{nn'})\right\rangle_C$.  We
compare $\sigma$ and $\sigma_u$ for weakly coupled planes with, e.g.
$W=1$ and $\gamma=0.9$ in Fig.~\ref{fig:CpE01+uniform}.  The peak
structure for this small disorder is a again finite size effect
reflecting the peaks of $\rho(E)$.  The positions of the minima of
$\sigma(E)$ are the same as expected from the joint density of states
but the minima are much more pronounced.  The reason for this behavior
is the strong localization of the states for energies with low
$\rho(E)$ similar to the localization in the band tails; the latter
causes the decrease of $\sigma$ at higher energies. Thus despite the
small system size the conductivity is highly influenced by
localization effects.

\begin{figure}[p]
  \centerline{\psfig{figure=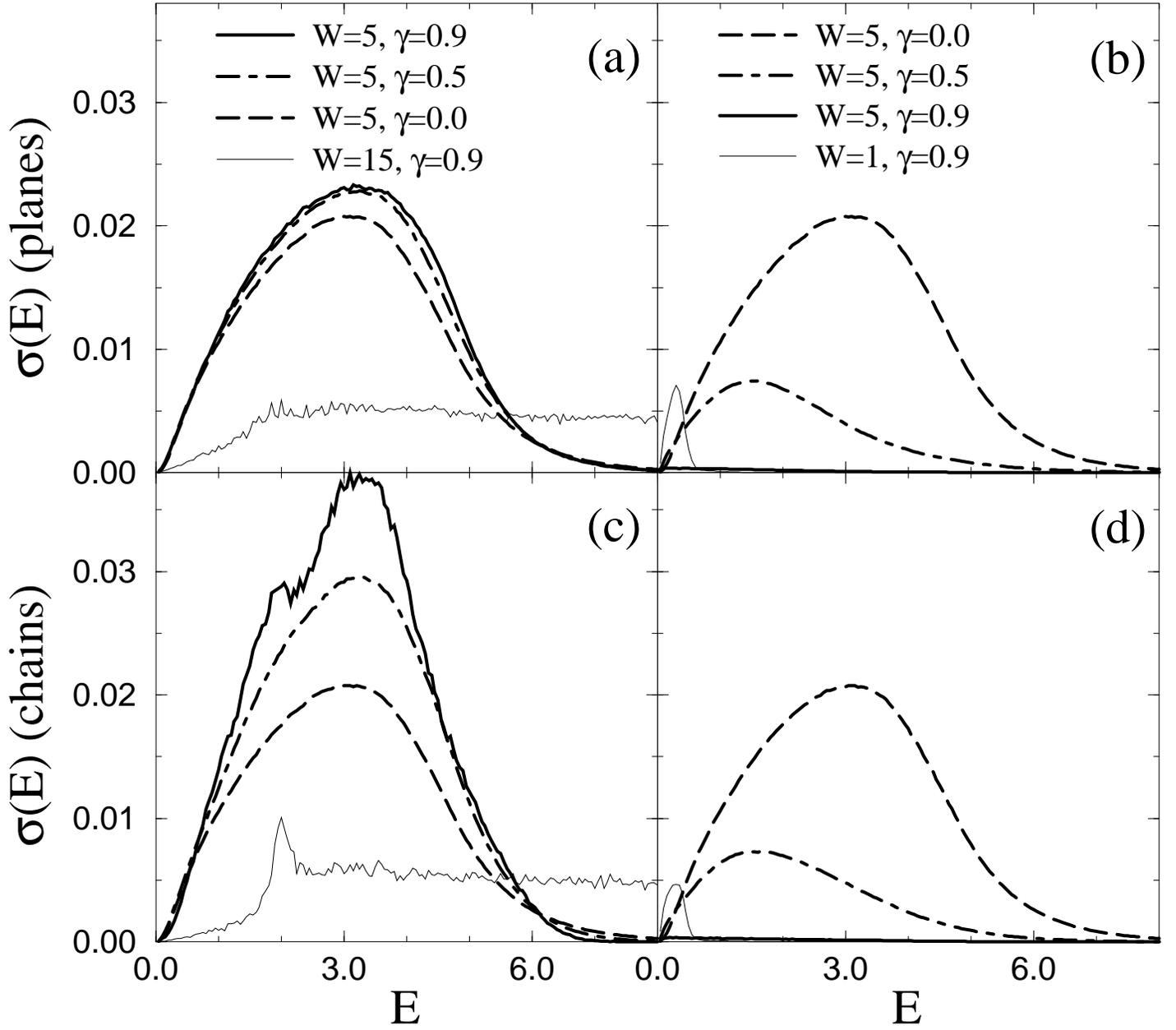}}
  \caption{Conductivity $\sigma$ for $N=12$ and various anisotropies
    $\gamma$ and disorders $W$ for weakly coupled chains (c), (d) and
    planes (a), (b) for the parallel (a), (c) and perpendicular (b),
    (d) direction.  The perpendicular conductivity for $\gamma=0.9$
    and $W=15$ is negligible while the parallel conductivity for
    $\gamma=0.9$ and $W=1$ exceeds the range of the diagram (see
    Fig.~\ref{fig:CpE01+uniform}).}
  \label{fig:L_gamma}
\end{figure}

In Fig.~\ref{fig:L_gamma} we present the conductivity computed for the
two nonequivalent directions: parallel (a), (c) and perpendicular (b),
(d) to the planes and chains, respectively. For strong anisotropy the
wave functions are concentrated to a few chains or planes as shown in
Fig.~\ref{fig:AWF01}. Consequently, the conductivity is drastically
reduced in the perpendicular direction. For $\gamma=0.9$ the maximum
of $\sigma$ is reached at small energies because the most extended
eigenstates appear in the band center. This causes the (small) peak
for $W=1$. For strong disorder $W=15$ all eigenstates are strongly
localized and the perpendicular conductivity is nearly negligible.
For the parallel conductivity we find an increase if the anisotropy
becomes stronger.  Here, the transport is not handicapped by the
anisotropic localization of the wave functions. The increase is
relatively small for the planes and considerable for the chains. A
good argument to explain this difference is the form of the density of
states which yields a higher amount of possibly transitions for the
energies around the position of the maximum of $\sigma(E)$ for the
second. The parallel conductivity at $W=15$ is relatively small but
considerable in a large energy range which reflects the disorder
widened energy band.  We note that the conductivities for a very
strong anisotropy $\gamma=0.99$ are nearly equal to those of
$\gamma=0.9$ in the parallel direction and again negligible in the
perpendicular direction.

%
%

\section{Conclusions}
\label{sec-concl}
In the present work we have studied the localization behavior of
eigenfunctions and transport properties of the Anderson model with
anisotropic hopping.  As expected from the general argument for the
fractal nature of wave functions at the metal-insulator transition,
multifractal eigenstates were found even for strong anisotropy. The
multifractal description holds not only directly at the transition but
also close to it due to the small system sizes considered. As a first
estimate for the critical disorder $W_c$ we determined that $W_{f_c}$
where the states show the characteristic singularity spectrum
$f_c(\alpha)$ which indicates the MIT in the isotropic case. But
especially for weakly coupled planes the computed anisotropy
dependence of the critical disorder differs remarkably from the TMM
results.  \cite{AEbenen2} We also analyzed the system-size dependences
of the singularity spectra to determine the MIT. The observed
$W_c(\gamma)$ agree reasonably well with the TMM
data.\cite{AEbenen2,AKetten} Therefore we conclude that the
''characteristic spectrum'' is no longer valid if the anisotropy
becomes strong. This is surprising because $f_c$ was independent of
the microscopic details of the isotropic system for the 3D case. The
spectrum at $W_c$ for weakly coupled planes is wider than $f_c$. This
coincides with the observed concentration of the probability amplitude
to only a few planes perpendicular to the direction with reduced
hopping for large anisotropy.

We have also studied the AC conductivity of small anisotropic samples
using Kubo's formula. In this case the treatable system sizes are very
small because all eigenvalues and eigenvectors of the Hamiltonian are
needed.  Nevertheless we observe a rapidly decreasing conductivity in
the direction with smaller hopping integral if the anisotropy becomes
stronger. This is a pure localization effect. For the used small
system size $N=12$ it is surprising that this can be observed, because
the characteristic length scales are much larger for nearly all of the
states. Another interesting fact is that the density of states for an
anisotropy $\gamma=0.9$ is already nearly identically with that of the
corresponding lower-dimensional system.

\acknowledgments This work was supported by the DFG within the
Sonderforschungsbereich 393. We thank A.\ Meyer and T.\ Apel for help
with parallelizing the Lanczos program.

%
%

\newpage

\newpage

\newpage

\newpage

\newpage

\newpage

\newpage

\newpage

\newpage

\end{document}